\documentstyle[12pt,psfig,graphics]{article}

\newcommand{\ds}{\displaystyle }
\newcommand{\R}{{\sf R\hspace*{-0.9ex}%
\rule{0.15ex}{1.5ex}\hspace*{0.9ex}}}
\newcommand{\Z}{{\sf Z\hspace*{-0.9ex}%
\rule{0.15ex}{1.5ex}\hspace*{0.9ex}}}

\title{DISPERSION RELATIONS AND WAVE OPERATORS IN SELF-SIMILAR QUASI-CONTINUOUS LINEAR CHAINS}

\author{{\sl\small T.M. Michelitsch $^{1}$\footnote{Corresponding author, Email:
michel@lmm.jussieu.fr } \, G.A. Maugin$^{1}$ F. C. G. A. Nicolleau$^2$, A. F. Nowakowski$^2$, S. Derogar$^3$} \\[5pt]
 $^1$ Institut Jean le Rond d'Alembert \\ CNRS UMR 7190 \\
 Universit\'{e} Pierre et Marie Curie, Paris 6 \\ FRANCE
 \\[5pt]
 \\
$^2$ Department of Mechanical Engineering\\
$^3$ Department of Civil and Structural Engineering\\
University of Sheffield\\
United Kingdom
\\[5pt]
 \\
{\it Physical Review E {\bf 80}, 011135 (2009)}}

\begin{document}

\maketitle

\section{Abstract}
{\small We construct self-similar functions and linear operators to deduce a
self-similar variant of the Laplacian operator and of the
D'Alembertian wave operator. The exigence of self-similarity as a
symmetry property requires the introduction of non-local
particle-particle interactions. We derive a self-similar linear wave
operator describing the dynamics of a quasi-continuous linear chain
of infinite length with a spatially self-similar distribution of
nonlocal inter-particle springs. The self-similarity of the nonlocal
harmonic particle-particle interactions results in a dispersion
relation of the form of a Weierstrass-Mandelbrot function which
exhibits self-similar and fractal features. We also derive a
continuum approximation which relates the self-similar Laplacian to
fractional integrals and yields in the low-frequency regime a power
law frequency-dependence of the oscillator density.
\newline\newline
\noindent {\bf Keywords:} Self-similarity, self-similar functions,
affine transformations, Weierstrass-Mandelbrot function, fractal
functions, fractals, power laws, fractional integrals.}
\newline\newline
\noindent {\bf PACS: 05.50.+q 81.05.Zx 63.20.D-}

\section{Introduction}
\label{intro}

In the seventies of the
last century the development of the {\it Fractal Geometry} by Mandelbrot \cite{1}
launched a scientific revolution. However, the mathematical roots of
this discipline originate much earlier
in the $19^{th}$ century \cite{8}.
The superior electromagnetic properties of ``{\it fractal antennae}"
have been known already for a while \cite{28,24}. More recently one found by means of numerical simulations that fractal gaskets such as the Sierpinski gasket reveal interesting vibrational properties
\cite{3}. Meanwhile physical
problems in fractal and self-similar structures or media become more and
more a subject of interest also in
analytical mechanics and engineering science. This is true in statics and dynamics.
However technological exploitations of effects based on self-similarity and ``fractality"
are still very limited due to a lack of fundamental understanding of the role of the self-similar symmetry.
An improved understanding could raise an enormous new field for
basic research and applications in a wide range of
disciplines including fluid mechanics and the mechanics of granular
media and solids. Some initial steps have been performed (see
papers \cite{3,2,tara,6,7,epstein} and the references therein).
However a generally accepted ``fractal mechanics" has yet to be
developed. Therefore, it is highly desirable to develop
sufficiently simple models which are on the one hand accessible to a mathematical-analytical framework
and on the other hand which capture the essential features imposed by self-similar scale invariant symmetry. The goal of this demonstration is to develop such a model.

Several significant contributions of
fractal and self-similar chains and lattices have been presented \cite{prl9122,prl3297,prl1239,prl3110}.
In these papers problems on {\it discrete} lattices
with fractal features are addressed. Closed form
solutions for the dynamic Green's function and the vibrational
spectrum of a linear chain with spatially exponential properties are
developed in a recent paper \cite{michelmaugin}.
A similar fractal type of linear chain as in the present paper has been considered
very recently by Tarasov \cite{tara}. Unlike in the present paper the chain considered by Tarasov in \cite{tara} is {\it discrete}, i.e.
there remains a characteristic length scale which is given by the next-neighbor distance of the particles.

In contrast to all these works we analyze in the present paper vibrational properties in a {\it quasi-continuous} linear chain with (in the self-similar limiting case) infinitesimal lattice spacing with a non-local spatially self-similar distribution of power-law-scaled harmonic inter-particle interactions (springs). In this way we avoid the appearance of a characteristic length scale in our chain model.
It seems there are analogue situations in turbulence \cite{weiermandel} and other areas where the present
interdisciplinary approach could be useful.

Our demonstration is organized as follows: \S~\ref{thomas1} is devoted
to the construction of self-similar functions and operators where a self-similar variant of the
Laplacian is deduced. This Laplacian gets his physical justification
in \S~\ref{thomas2}.
It is further shown in \S~\ref{thomas1} that in a continuum approximation this Laplacian takes the form of fractional integrals.
In \S~\ref{thomas2} we consider a self-similar quasi-continuous linear chain with self-similar harmonic
interactions. The equation of motion of this chain takes the form of a self-similar wave equation containing the self-similar Laplacian defined in \S~\ref{thomas1} leading to a dispersion relation having the form of the Weierstrass-Mandelbrot function which is a self-similar and for a certain parameter range also a fractal function.

\section{Construction of self-similar functions and linear operators}
\label{thomas1}

In this paragraph we define the term ``self-similarity" with respect
to functions and operators. We call a scalar function $\phi(h)$ {\it
exact self-similar} with respect to variable $h$ if the condition
\begin{equation}
\label{self-sim}
\phi(Nh)=\Lambda\phi(h)
\end{equation}
is satisfied for all values $h>0$ of the scalar variable $h$.
We call (\ref{self-sim}) the ``affine
problem"\footnote{where we restrict here to affine transformations
$h'=Nh+c$ with $c=0$.} where $N$ is a fixed parameter and $\Lambda=N^{\delta}$
represents a continuous set of admissible eigenvalues.
The band of admissible $\delta=\frac{\ln{\Lambda}}{\ln{N}}$ is to be determined.
A function $\phi(h)$ satisfying (\ref{self-sim})
for a certain $N$ and admissible $\Lambda=N^{\delta}$ represents an unknown ``solution" to the affine problem of the form $\phi_{N,\delta}(h)$ and is to be determined.

As we will see
below for a given $N$ solutions $\phi(h)$ {\it exist} only in a certain
range of admissible $\Lambda$. From the definition of the problem
follows that if $\phi(h)$ is a solution of (\ref{self-sim}) it is
also a solution of $\phi(N^sh)=\Lambda^s\phi(h)$ where $s\in \Z$ is discrete and can
take all positive and negative integers including zero. We
emphasize that non-integer $s$ are not admitted. The discrete set of
pairs $\Lambda^s,N^s$ are for all $s \in \Z$ related by a power law
with the same power $\delta$, i.e. $\Lambda=N^{\delta}$ hence
$\Lambda^s=(N^{s})^{\delta}$.
By replacing $\Lambda$ and $N$ by $\Lambda^{-1}$ and $N^{-1}$ in (\ref{self-sim})
defines the identical problem. Hence we can restrict our considerations on fixed values of
$N>1$.

We can consider the affine problem (\ref{self-sim}) as the eigenvalue problem for a
linear operator ${\hat A}_{N}$ with a certain given fixed parameter
$N$ and eigenfunctions $\phi(h)$ to be determined which correspond
to an {\it admissible} range of eigenvalues $\Lambda=N^{\delta}$ (or
equivalently to an admissible range of exponent
$\delta={\ln{\Lambda}}/{\ln {N}}$). For a function $f(x,h)$ we
denote by ${\hat A}_{N}(h)f(x,h)=:f(x,Nh)$ when the affine
transformation is only performed with respect to variable $h$.

We assume $\Lambda, N \in \R$ for physical reasons without too much
loss of generality to be real and positive. For our convenience we define the
``affine" operator ${\hat A}_{N}$ as follows

\begin{equation}
\label{affinetrafo} {\hat A}_{N}f(h)=: f(Nh)
\end{equation}
It is easily verified that the affine operator ${\hat A}_{N}$ is
{\it linear}, i.e.
\begin{equation}
\label{linearite} {\hat
A}_{N}\left(c_1f_1(h)+c_2f_2(h)\right)=c_1f_1(Nh)+c_2f_2(Nh)
\end{equation}
and
\begin{equation}
\label{pouvoir} {\hat A}_{N}^sf(h)=f(N^sh),\hspace{2cm}
s=0\pm1,\pm2, ..\pm\infty
\end{equation}

We can define affine operator functions for
any smooth function $g(\tau)$ that can be expanded into a Maclaurin
series as

\begin{equation}
\label{Taylor} g(\tau)=\sum_{s=0}^{\infty}a_s\tau^s
\end{equation}
We define an affine operator function in the form
\begin{equation}
\label{fonctionop} g(\xi{\hat A}_N)=\sum_{s=0}^{\infty}a_s\xi^s{\hat
A}_N^s
\end{equation}
where $\xi$ denotes a scalar parameter. The operator function which
is defined by (\ref{fonctionop}) acts on a function $f(h)$ as
follows
\begin{equation}
\label{fonctionopact} g(\xi{\hat
A}_N)f(h)=\sum_{s=0}^{\infty}a_s\xi^sf(N^sh)
\end{equation}
where relation (\ref{pouvoir}) with expansion (\ref{fonctionop}) has
been used. The convergence of series (\ref{fonctionopact}) has to be
verified for a function $f(h)$ to be admissible. An explicit
representation of the affine operator ${\hat A}_N$ can be obtained
when we write $f(h)=f(e^{\ln h})={\bar f}(\ln h)$ to arrive at

\begin{equation}
\label{affineexplicit} {\hat A}_N(h)=e^{\ln N\frac{\rm d}{\rm d (\ln
h)}}
\end{equation}

This relation is immediately verified in view of
\begin{equation}
{\hat A}_N(h)f(h)=e^{\ln N\frac{\rm d}{\rm d (\ln h)}}f(e^{\ln
h})=f(e^{\ln h +\ln N})=f(Nh)
\end{equation}

With this machinery we are now able to construct self-similar
functions and operators.

\subsection{Construction of self-similar functions}

A self-similar function solving problem (\ref{self-sim}) is formally
given by the series
\begin{equation}
\label{self-sim-func}
\phi(h)=\sum_{s=-\infty}^{\infty}\Lambda^{-s}{\hat A}_{N}^sf(h)=\sum_{s=-\infty}^{\infty}\Lambda^{-s}f(N^sh)
\end{equation}
for any function $f(h)$ for which the series (\ref{self-sim-func})
is uniformly convergent for all $h$. We introduce
the self-similar operator

\begin{equation}
\label{selfsimop}
{\hat
T}_N=\sum_{s=-\infty}^{\infty}\Lambda^{-s}{\hat A}_{N}^s
\end{equation}
that fulfils formally the condition of self-similarity ${\hat
A}_N{\hat T}_N=\Lambda {\hat T}_N$ and hence (\ref{self-sim-func})
solves the affine problem (\ref{self-sim}). In view of the symmetry
with respect to inversion of the sign of $s$ in
(\ref{self-sim-func}) and (\ref{selfsimop}) we can restrict
ourselves to $N>1$ ($N,\Lambda \in \R$) without any loss of
generality\footnote{We also can exclude the trivial case $N=1$.}:
Let us look for admissible functions $f(t)$ for which
(\ref{self-sim-func}) is convergent. To this end we have to demand
simultaneous convergence of the partial sums over positive and
negative $s$. Let us assume that (where we can confine ourselves to
$t>0$)
\begin{equation}
\label{ass1} \lim_{t\rightarrow 0} f(t)=a_0\,t^{\alpha}
\end{equation}
For $t \rightarrow \infty$ we have to demand that $|f(t)|$ increases
not stronger than a power of $t$, i.e.
\begin{equation}
\label{ass2} \lim_{t\rightarrow \infty} f(t)=c_{\infty}\, t^{\beta}
\end{equation}
with $a_0, c_{\infty}$ denoting constants. Both exponents
$\alpha,\beta \in \R$ are allowed to take positive or negative
values which do not need to be integers. A brief consideration of
partial sums yields the following requirements for
$\Lambda=N^{\delta}$, namely: Summation over $s<0$ in
(\ref{self-sim-func}) requires absolute convergence of a geometrical
series leading to the condition for its argument $\Lambda
N^{-\alpha}<1$. That is we have to demand $\delta<\alpha$. The
partial sum over $s>0$ requires absolute convergence of a
geometrical series leading to the condition for its argument
$\Lambda^{-1} N^{\beta}< 1$ which corresponds to $\delta > \beta$.
Both conditions are simultaneously met if
\begin{equation}
\label{lam}
N^{\beta} < \Lambda=N^{\delta} < N^{\alpha}
\end{equation}
or equivalently
\begin{equation}
\label{lam2}
\beta < \delta=\frac{\ln{\Lambda}}{\ln{N}} < \alpha
\end{equation}

Relations (\ref{lam}) and (\ref{lam2}) require additionally $\beta <
\alpha$, that is only functions $f(t)$ with the behaviour
(\ref{ass1}) and (\ref{ass2}) with $\beta < \alpha$ are {\it
admissible} in (\ref{self-sim-func}). The case $\beta=0$ includes for instance certain bounded functions
$|f(t)| < M$ such as some periodic functions.

\subsection{A self-similar analogue to the Laplace operator}

In the sprit of (\ref{self-sim-func}) and (\ref{selfsimop}) we
construct an exactly self-similar function from the second
difference according to

\begin{equation}
\label{uhdef}
\phi(x,h)=
{\hat T}_N(h)\left(u(x+h)+u(x-h)-2u(x)\right)
\end{equation}
where $u(..)$ denotes an arbitrary smooth continuous field variable and
${\hat T}_N(h)$ expresses that the affine operator ${\hat
A}_N(h)$ acts only on the dependence on $h$, that is ${\hat
A}_N(h)v(x,h)=v(x,Nh)$. We have with $\xi=\Lambda^{-1}$ the
expression
\begin{equation}
\label{uh}
\phi(x,h)=
\sum_{s=-\infty}^{\infty}\xi^s\left\{u(x+N^sh)+u(x-N^sh)-2u(x)\right\}
\end{equation}
which is a self-similar function with respect to its dependence on
$h$ with ${\hat A}_N(h)\phi(x,h)$ $= \phi(x,Nh)=\xi^{-1}\phi(x,h)$
but a regular function with respect to $x$. The function $\phi(x,h)$
exists if the series (\ref{uh}) is convergent. Let us assume that
$u(x)$ is a smooth function with a convergent Taylor series for any
$h$. Then we have with $u(x\pm h)=e^{\pm h\frac{d}{dx}}u(x)$ and
$u(x+h)+u(x-h)-2u(x)=\left(
e^{h\frac{d}{dx}}+e^{-h\frac{d}{dx}}-2\right)u(x)$ which can be
written as
\begin{equation}
\label{diffeq} u(x+h)+u(x-h)-2u(x)=
4\sinh^2\left({\frac{h}{2}\frac{d}{dx}}\right)u(x)=h^2\frac{d^2}{dx^2}u(x)+orders\,\,{h^{\geq
4}}
\end{equation}
thus $\alpha=2$ in criteria (\ref{ass1}) is met. If we demand
$u(x)$ being Fourier transformable we have as necessary condition
that
\begin{equation}
\label{uint} \int_{-\infty}^{\infty}|u(x)|\,{\rm d}x < \infty
\end{equation}
exists. This is true if $|u(t)|$ tends to zero as $t\rightarrow\pm
\infty$ as $|t|^{\beta}$ where $\beta <-1$. We have then the
condition that
\begin{equation}
\label{expon}
\beta < 0 <  \delta=- \frac{\ln{\xi}}{\ln{N}} < \alpha = 2
\end{equation}
We will see below that only $\delta>0$ is {\it physically
admissible}, i.e. compatible with harmonic particle-particle
interactions which decrease with increasing particle-particle
distance.

The 1D Laplacian $\Delta_1$ is defined by
\begin{equation}
\label{1DLaplacian}
\Delta_1 u(x)= \frac{d^2}{dx^2}u(x)= \lim_{\tau \rightarrow 0}\frac{\left(u(x+\tau)+u(x-\tau)-2u(x)\right)}{\tau^2}
\end{equation}
Let us now define a self-similar analogue to the 1D Laplacian (\ref{1DLaplacian})
where we put with $\xi=N^{-\delta}$
\begin{eqnarray}
\label{Laplautos} \Delta_{(\delta,N,\tau)}u(x) =: const
\lim_{\tau\rightarrow 0} \tau^{-\lambda}
\phi(x,\tau)
\\
 =
const
\,\, \lim_{\tau\rightarrow
0} 
{
\tau^{-\lambda}
\sum_{s=-\infty}^{\infty}}\xi^s\left(u(x+N^s\tau)+u(x-N^s\tau)-2u(x)\right)
\end{eqnarray}
where we have introduced a renormalisation-multiplier
$\tau^{-\lambda}$ with the unknown power $\lambda$ to be determined such that the limiting case is finite. The constant factor
$const$ indicates that there is a certain arbitrariness in this
definition and will be chosen conveniently. Let us consider the
limit $\tau\rightarrow 0$ by the special sequence $\tau_n=N^{-n}h$
with $n\rightarrow\infty$ and $h$ being constant. Unlike in the 1D
case (\ref{1DLaplacian}), the result of this limiting process
depends crucially on the choice of the sequence $\tau_n$. We see here that
the self-similar Laplacian cannot be defined uniquely as in the 1D case.
We have (by putting in (\ref{Laplautos}) $
const=h^{\lambda}$)
\begin{equation}
\label{Laplaself}
\Delta_{(\delta,N,h)}u(x)= 
\lim_{n\rightarrow \infty}
 {
 N^{\lambda n} \xi^n\sum_{s=-\infty}^{\infty}\xi^{s-n}\left(u(x+N^{s-n}h)+u(x-N^{s-n}h)-2u(x)\right)}
\end{equation}
which assumes by replacing $s-n\rightarrow s$ the form
\begin{equation}
\label{Laplaself2}
\Delta_{(\delta,N,h)}u(x)=\phi(x,h)\lim_{n\rightarrow
\infty}N^{-(\delta-\lambda) n}
\end{equation}
which is only finite and nonzero if $\lambda=\delta$. The
``Laplacian" can then be defined simply by
\begin{equation}
\label{Laplaself3}
 \Delta_{(\delta,N,h)}u(x)=: \lim_{n\rightarrow \infty}N^{\delta
 n}\phi(x,N^{-n}h) =\phi(x,h)
 \end{equation}
or by using (\ref{uhdef}) and (\ref{diffeq}) we can simply
write\footnote{We have to replace $\frac{d}{dx} \rightarrow
\frac{\partial }{\partial x}$ if the Laplacian acts on a field
$u(x,t)$ as in Sec. \ref{thomas2}.}

\begin{equation}
\label{lapla}
 \Delta_{(\delta,N,h)}=4{\hat
 T}_N(h)\sinh^2\left({\frac{h}{2}\frac{\partial}{\partial
 x}}\right)=4\sum_{s=-\infty}^{\infty}N^{-\delta s}\sinh^2\left({\frac{N^sh}{2}\frac{\partial}{\partial
 x}}\right)
\end{equation}
where ${\hat T}_N(h)$ is the self-similar operator defined in
(\ref{selfsimop}). The self-similar analogue of Laplace operator
defined by (\ref{lapla}) depends on the parameters $\delta,N,h$. We
furthermore observe the self-similarity of Laplacian (\ref{lapla}) with respect to its dependence on $h$,
namely

\begin{equation}
\label{selflap}
\Delta_{(\delta,N,Nh)}=N^{\delta}\Delta_{(\delta,N,h)}
\end{equation}

\subsection{Continuum approximation - link to fractional integrals}
\label{contapprox}

For numerical evaluations it may be convenient to utilize a
continuum approximation of the self-similar Laplacian (\ref{lapla}).
To this end we put $N=1+\epsilon$ (with $0<\epsilon \ll 1$ thus
$\epsilon\approx\ln{N}$) where $\epsilon$ is assumed to be ``small"
and $s\epsilon=v$ such that ${\rm d}v\approx \epsilon$ and
$N^s=(1+\epsilon)^{\frac{v}{\epsilon}}\approx e^v$. In this
approximation $N^s\approx e^v$ becomes a (quasi)-continuous variable
when $s$ runs through $s\in \Z$. Then we can write
(\ref{self-sim-func}) in the form

\begin{equation}
\label{Laplacesum}\phi(h)=\sum_{s=-\infty}^{\infty}N^{-s\delta}f(N^sh)\approx
\frac{1}{\epsilon}\int_{-\infty}^{\infty}e^{-\delta v}f(h e^v){\rm d}v
\end{equation}
which can be further written with $he^{v}=\tau$ ($h>0$) and
$\frac{{\rm d}\tau}{\tau}={\rm d}v$ and $\tau(v\rightarrow
-\infty)=0$ and $\tau(v\rightarrow \infty)=\infty$ as
\begin{equation}
\label{fract} \phi(h)\approx
\frac{h^{\delta}}{\epsilon}\int_0^{\infty}\frac{f(\tau)}{\tau^{1+\delta}}\,{\rm
d}\tau
\end{equation}
In this continuum approximation the function $\phi(h)$ obeys the
same scaling behaviour as (\ref{self-sim-func}), namely
$\phi(h\lambda)=\lambda^{\delta}\phi(h)$ but in contrast to
(\ref{self-sim-func}) $\lambda$ can assume any continuous positive
value. This is due to the fact that (\ref{fract}) is holding for
$N=1+\epsilon$ with sufficiently small $\epsilon>0$ since in this
limiting case there exists for any continuous value $\lambda>0$ an
$m \in \Z$ such that $N^m\approx\lambda$. The existence requirement
for integral (\ref{fract}) leads to the same requirements for $f(t)$
as in (\ref{self-sim-func}), namely inequality (\ref{lam2}).
Application of the approximate relation (\ref{fract}) to Laplacian
(\ref{lapla}) yields
\begin{equation}
\label{fractlapself} \Delta_{(\delta,\epsilon,h)}u(x)\approx
\frac{h^{\delta}}{\epsilon}\int_0^{\infty}\frac{(u(x-\tau)+u(x+\tau)-2u(x))}{\tau^{1+\delta}}\,{\rm
d}\tau
\end{equation}
where this integral exists for $\beta < 0 < \delta < 2$ and $\beta < -1$
because the required existence of integral (\ref{uint}) and relation
(\ref{diffeq}).
By performing two partial integrations and by taking into account
the vanishing boundary terms at $\tau=0$ and $\tau=\infty$ for
$0 < \delta < 2$, we can re-write (\ref{fractlapself}) in the form
of a convolution of the conventional 1D Laplacian $\frac{d^2 u}{
dx^2}(x)$, namely

\begin{equation}
\label{reforme} \Delta_{(\delta,\epsilon,h)}u(x)\approx
\int_{-\infty}^{\infty}g(|x-\tau|)\frac{d^2 u}{ d\tau^2}(\tau){\rm
d}\tau
\end{equation}
with the kernel

\begin{equation}
\label{kernel} \ds
g(|x|)=\frac{h^{\delta}}{\delta(\delta-1)\epsilon}\,
|x|^{1-\delta} , \hspace{1cm} \delta \neq 1
\end{equation}
where $0 < \delta < 2$ and $g(|x|)=-\frac{h}{\epsilon}\ln{|x|}$ for $\delta=1$.
We can further write for $\delta \neq 1$
(\ref{reforme}) in terms of {\it fractional integrals}

\begin{equation}
\label{fractionalaplaican} \ds
\Delta_{(\delta=2-D,\epsilon,h)}u(x)\approx \frac{h^{2-D}}{\epsilon}
\frac{\Gamma(D)}{(D-1)(D-2)}\left({\cal
D}_{-\infty,x}^{-D}+(-1)^D{\cal
D}_{\infty,x}^{-D}\right)\Delta_1u(x)
\end{equation}
where $\Delta_1u(x)=\frac{d^2}{dx^2}u(x)$ denotes the conventional
1D-Laplacian and $D=2-\delta>0$ which is positive in the admissible range
of $0<\delta<2$. For $0<\delta<1$ the quantity $D$ can be identified
with the estimated fractal dimension of the fractal dispersion
relation of the Laplacian \cite{Hardy} which is deduced in the next
section. In (\ref{fractionalaplaican}) we have introduced the
Riemann-Liouville fractional integral ${\cal
D}_{a,x}^{-D}$ which is defined by(e.g. \cite{miller,kilbas})

\begin{equation}
\label{fractional} {\cal
D}_{a,x}^{-D}v(x)=\frac{1}{\Gamma(D)}\int_a^x(x-\tau)^{D-1}v(\tau){\rm
d}\tau
\end{equation}
where ${\Gamma(D)}$ denotes the $\Gamma$-function which represents
the generalization of the factorial function to non-integer $D>0$.
The $\Gamma$-function is defined as

\begin{equation}
\label{gamma} \Gamma(D)=\int_0^{\infty}\tau^{D-1} e^{-\tau} {\rm d}
\tau \,,\hspace{1cm} D > 0
\end{equation}
For positive integers $D>0$ the $\Gamma$-function reproduces the
factorial-function $\Gamma(D)=(D-1)!$ with $D=1,2,..\infty$.
\newline\newline

\section{The physical chain model}
\label{thomas2}

We consider an infinitely long quasi-continuous linear chain of
identical particles. Any space-point $x$ corresponds to a ``material
point" or particle. The mass density of particles is assumed to be
spatially homogeneous and equal to one for any space point $x$. Any
particle is associated with one degree of freedom which is
represented by the displacement field $u(x,t)$ where $x$ is its
spatial (Lagrangian) coordinate and $t$ indicates time. In this
sense we consider a quasi continuous spatial distribution of
particles. Any particle at space-point $x$ is non-locally connected
by harmonic springs of strength $\xi^s$ to particles located at
$x\pm N^s h$, where $N>1$ and $N \in \R$ is not necessarily integer,
$h>0$, and $s=0, \pm 1,\pm 2,..\pm\infty $. The requirement of
decreasing spring constants with increasing particle-particle
distance leads to the requirement that $\xi=N^{-\delta} < 1$ ($N>1$)
i.e. only chains with $\delta > 0$ are physically admissible. In
order to get exact self-similarity we avoid the notion of
``next-neighbour particles" in our chain which would be equivalent
to the introduction of an internal length scale (the next neighbour
distance). To admit particle interactions over arbitrarily close
distances $N^s h\rightarrow 0$ ($s\rightarrow -\infty$, $h=const$)
our chain has to be {\it quasi-continuous}. This is the principal
difference to the {\it discrete} chain considered recently by
Tarasov \cite{tara} which is discrete and not self-similar.

The Hamiltonian which describes our chain can be written as

\begin{equation}
\label{Hamiltonian}
H=
\frac{1}{2}\int_{-\infty}^{\infty}\left(\dot{u}^2(x,t)+{\cal V}(x,t,h)\right){\rm d}x
\end{equation}
In the spirit of (\ref{self-sim-func}) the elastic energy density
${\cal V}(x,t,h)$ is assumed to be constructed self-similarly,
namely\footnote{The additional factor ${1}/{2}$ in the elastic
energy avoids double counting.}

\begin{equation}
\label{homchain} {\cal V}(x,t,h)=\frac{1}{2}{\hat
T}_N(h)\left[(u(x,t)-u(x+h,t))^2 +(u(x,t)-u(x-h,t))^2\right]
\end{equation}
where ${\hat T}_N(h)$ is the self-similar operator (\ref{selfsimop})
with $\xi=\Lambda^{-1}=N^{-\delta}$ to arrive at

\begin{equation}
\label{elastic}
{\cal V}(x,t,h)=\frac{1}{2}\sum_{s=-\infty}^{\infty}\xi^s\left[(u(x,t)-u(x+hN^s,t))^2
+(u(x,t)-u(x-hN^s,t))^2\right]
\end{equation}
The elastic energy density ${\cal V}(x,t,h)$ fulfills the
condition of self-similarity with respect to $h$, namely
\begin{equation}
\label{selfsimelasr}
{\hat A}_N(h){\cal V}(x,t,h)={\cal V}(x,t,Nh)= \xi^{-1}{\cal V}(x,t,h)
\end{equation}

We have to demand in our physical model that the energy is finite,
i.e. (\ref{elastic}) needs to be convergent which yields $\alpha=2$ as
for the Laplacian (\ref{uh}). To determine $\beta$ we have to demand
that $u(x,t)$ is a Fourier transformable field\footnote{This assumption
defines the (function) space of eigenmodes and corresponds to
infinite body boundary conditions.}. Thus we have to have an
asymptotic behaviour of $|u(x\pm \tau,t)| \rightarrow 0$ as $\tau^{\beta}$
where $\beta <-1$ as $\tau\rightarrow \infty$. From this follows
$|u(x,t)-u(x\pm \tau,t)|^2$ behaves then as $|u(x,t)|^2$.
Hence, the elastic energy density (\ref{elastic})
is finite if
\begin{equation}
\label{interval2}
 0 < \delta < 2
\end{equation}
where $\beta < -1$.

This inequality determines the range
of the admissible values of $\delta$ in order to achieve
convergence.
The equation of motion is obtained from
\begin{equation}
\label{eqmo} \frac{\partial^2 u}{\partial t^2}=-\frac{\delta
H}{\delta u}
\end{equation}
(where $\delta . / \delta u$ stands for a functional derivative) to arrive at
\begin{eqnarray}
\frac{\partial^2 u}{\partial t^2} &=&
- \sum_{s=-\infty}^{\infty} \xi^s \left( 2u(x,t)-u(x+hN^s,t)-u(x-hN^s,t) \right)
\label{eqmo2a}
\\
\frac{\partial^2 u}{\partial t^2} &=& \Delta_{(\delta,N,h)}u(x,t)
\label{eqmo2}
\end{eqnarray}
with the self-similar Laplacian $\Delta_{(\delta,N,h)}$ of equation (\ref{lapla}).
As the elastic energy density (\ref{elastic})
the equation of motion is convergent for $\delta$ being in the interval (\ref{interval2})
where $\beta < -1$. We can re-write (\ref{eqmo2}) in the compact
form of a wave equation

\begin{equation}
\label{comp} \Box_{\small(\delta,N,h)}u(x,t)=0
\end{equation}
where $\Box_{\small(\delta,N,h)}$ is the {\it self-similar analogue
of the d'Alembertian wave operator} having the form

\begin{equation}
\label{comp2}
\Box_{\small(\delta,N,h)}=\Delta_{(\delta,N,h)}-\frac{\partial^2}{\partial
t^2}
\end{equation}
The d'Alembertian (\ref{comp2}) with the Laplacian (\ref{lapla})
describes the wave propagation in the self-similar chain with
Hamiltonian (\ref{Hamiltonian}). The present approach seems to be useful
as a point of departure to establish a generalized theory of wave propagation in self-similar
media.

Now we determine the dispersion relation, which is
constituted by the (negative) eigenvalues of the (semi-)negative
definite Laplacian (\ref{lapla}). To this end we make use of the
fact that the displacement field $u(x,t)$ is Fourier transformable
(guaranteed by choosing $\beta < -1$) and that
the exponentials $e^{ikx}$ are eigenfunctions of the self-similar
Laplacian (\ref{lapla}). We hence write the Fourier integral
\begin{equation}
\label{ufour} u(x,t)=\frac{1}{2\pi}\int_{-\infty}^{\infty}{\tilde
u}(k,t)e^{ikx}{\rm d}k
\end{equation}
to re-write (\ref{eqmo2}) for the Fourier amplitudes ${\tilde
u}(k,t)$ in the form

\begin{equation}
\label{eqmofou} 
\frac{\partial^2 \tilde{u}}{\partial t^2}(k,t)=-{\bar
\omega}^2(k)\,{\tilde u}(k,t)
\end{equation}
and obtain

\begin{equation}
\label{mandelbrot}{
\omega^2(kh)= 4\sum_{s=-\infty}^{\infty}N^{-\delta
s}\sin^2(\frac{khN^s}{2})}
\end{equation}

The series (\ref{mandelbrot}) describes a {\it Weierstrass-Mandelbrot
function} which is a continuous and for $0<\delta \leq 1$ a nowhere differentiable function
\cite{1,Hardy}. The Weierstrass-Mandelbrot function (\ref{mandelbrot}) fulfills the condition of self-similar symmetry, namely
\begin{equation}
\label{selfgrap} \omega^2(Nkh)=N^{\delta}\,\omega^2(kh)
\end{equation}
where the interval of convergence of the series of the {\it Weierstrass-Mandelbrot
function}
(\ref{mandelbrot}) is also given by (\ref{interval2}).
We emphasize that indeed {\it only} exponents $\delta$ in the
interval (\ref{interval2}) are {\it admissible} in Hamiltonian
(\ref{Hamiltonian}) with the elastic energy density (\ref{elastic})
in order to have a ``well-posed" problem.

It was shown by Hardy \cite{Hardy} that for $\xi N>1$ and
$\xi=N^{-\delta} <1$ or equivalently for
\begin{equation}
\label{deltaconv2} 0 < \delta < 1
\end{equation}
the Weierstrass-Mandelbrot function of the form (\ref{mandelbrot})
is not only self-similar but also a {\it fractal} curve of
(estimated) non-integer fractal (Hausdorff) dimension $D=2-\delta >
1$. Figs. 2-4 show dispersion curves $\omega^2(kh)$ for different
decreasing values of admissible $0< \delta < 1$ and increasing
fractal dimension $D$. Fig. 1 corresponds to the non-fractal case ($\delta=1.2>1$).
The increase of the fractal dimension from Figs.
2-4 is indicated by the increasingly irregular harsh behaviour of
the curves. In Fig. 4 the fractal dimension of the dispersion curve is with $D=1.9$ already close to the plane-filling dimension $2$.

To evaluate (\ref{mandelbrot}) approximately it is convenient to
replace the series by an integral utilizing a similar substitution
as in Sec. \ref{contapprox} ($\epsilon\approx \ln N$). By doing so
we smoothen the Weierstrass-Mandelbrot function (\ref{mandelbrot}).
It is important to notice that the resulting approximate dispersion
relation is hence differentiable and has not any more a fractal
dimension $D>1$ in the interval (\ref{deltaconv2}). For sufficiently
``small" $|k|h$ ($h>0$), i.e. in the long-wave regime we arrive at

\begin{equation}
\label{contapp} \omega^2(kh)\approx
\frac{(h|k|)^{\delta}}{\epsilon}C
\end{equation}
which is only finite if $(|k|h)^{\delta}$ is in the order of
magnitude of $\epsilon$ or smaller. This regime which includes the long-wave
limit $k \rightarrow 0$ is hence characterized by a power law
behaviour ${\bar \omega}(k)\approx Const\, |k|^{\delta/2}$ of the
dispersion relation. The constant $C$ introduced in (\ref{contapp})
is given by the integral

\begin{equation}
\label{CCabrev}
C=2\int_0^{\infty}\frac{(1-\cos{\tau})}{\tau^{1+\delta}}{\rm d}\tau
\end{equation}
which exists for $\delta$ being within interval (\ref{interval2}).

This approximation holds for ``small" $\epsilon \approx \ln N \neq
0$ ($0< \epsilon \ll 1$)\footnote{$\epsilon=0$ has to be excluded
since it corresponds to $N=1$.} which corresponds to the limiting
case that $N^s=e^v$ is continuous. In this limiting case we obtain
the oscillator density from \cite{michelmaugin}\footnote{The
additional prefactor ''$2$" takes into account the two branches of
the dispersion relation (\ref{mandelbrot}) (one for $k<0$ and one
for $k>0$).}

\begin{equation}
\label{oscden} \rho(\omega)= 2\frac{1}{2\pi}\frac{d |k|}{d\omega}
\end{equation}
which is normalized such that $\rho(\omega){\rm d}\omega$ counts the
number (per unit length) of normal oscillators having frequencies
within the interval $[\omega,\omega+{\rm d}\omega]$. We obtain then a power law of the form

\begin{equation}
\label{oscden2} \rho(\omega)= \ds \frac{2}{\pi\delta
h}\left(\frac{\epsilon}{C}\right)^{\frac{1}{\delta}}\omega^{\frac{2}{\delta}-1}
\end{equation}
where $\delta$ is restricted within interval (\ref{interval2}). We observe hence that
the power $\frac{2}{\delta}-1$ is restricted within the range $0< \frac{2}{\delta}-1 < \infty$
for $0<\delta<2$, especially with always vanishing oscillator density $\rho(\omega\rightarrow 0)=0$.

We emphasize that neither is the dependence on $k$ of the
Weierstrass-Mandelbrot function (\ref{mandelbrot}) represented by a
{\it continuous} $|k|^{\delta}$-dependence nor is this function
differentiable with respect to $k$. Application of (\ref{oscden}) is
hence only justified to be applied to the approximative
representation (\ref{contapp}) if $0 < \epsilon \ll 1$ thus
$N=1+\epsilon$ is sufficiently close to $1$ so that $N^s$ is a
quasi-continuous function when $s$ runs through $s\in \Z$. Hence
(\ref{oscden}) is not generally applicable to (\ref{mandelbrot}) for
any arbitrary $N>1$. We can consider (\ref{oscden2}) as the
low-frequency regime $\omega\rightarrow 0$ of the oscillator density
holding {\it only} in the quasi-continuous case $N=1+\epsilon$ with
$0<\epsilon \ll 1$.

\section{Conclusions}

We have depicted how self-similar functions and linear operators can
be constructed in a simple manner by utilizing a certain category of
conventional ``admissible" functions. This approach enables us to
construct non-local self-similar analogues to the Laplacian and
d'Alembert wave operator. The linear self-similar equation of motion
describes the propagation of waves in a quasi-continuous linear
chain with harmonic non-local self-similar particle-interactions.
The complexity which comes into play by the self-similarity of the
non-local interactions is completely captured by the dispersion
relations which assume the forms of Weierstrass-Mandelbrot functions
(\ref{mandelbrot}) exhibiting exact self-similarity and for certain
parameter combinations (relation (\ref{deltaconv2})) fractal
features. In a continuum approximation the self-similar Laplacian is
expressed in terms of fractional integrals (eq.
(\ref{fractionalaplaican})) leading for small $k$ (long-wave limit)
to a power-law dispersion relation (eq. (\ref{contapp})) and to a
power-law oscillator density (eq. (\ref{oscden2})) in the
low-frequency regime.

The self-similar wave operator (\ref{comp2}) with the Laplacian
(\ref{lapla}) can be generalized to describe wave propagation in
fractal and self-similar structures which are fractal subspaces
embedded in Euclidean spaces of 1-3 dimensions. The development of
such an approach could be a crucial step towards a better
understanding of the dynamics in materials with scale hierarchies of
internal structures (``multiscale materials") which may be idealized
as fractal and self-similar materials.

We hope to inspire further work and collaborations in this direction
to develop appropriate approaches useful for the modelling of static
and dynamic problems in self-similar and fractal structures in a
wider interdisciplinary context.

\section{Acknowledgements}
Fruitful discussions with J.-M. Conoir, D. Queiros-Conde and A.
Wunderlin are gratefully acknowledged.

\begin{figure}[t]
\psfig{figure=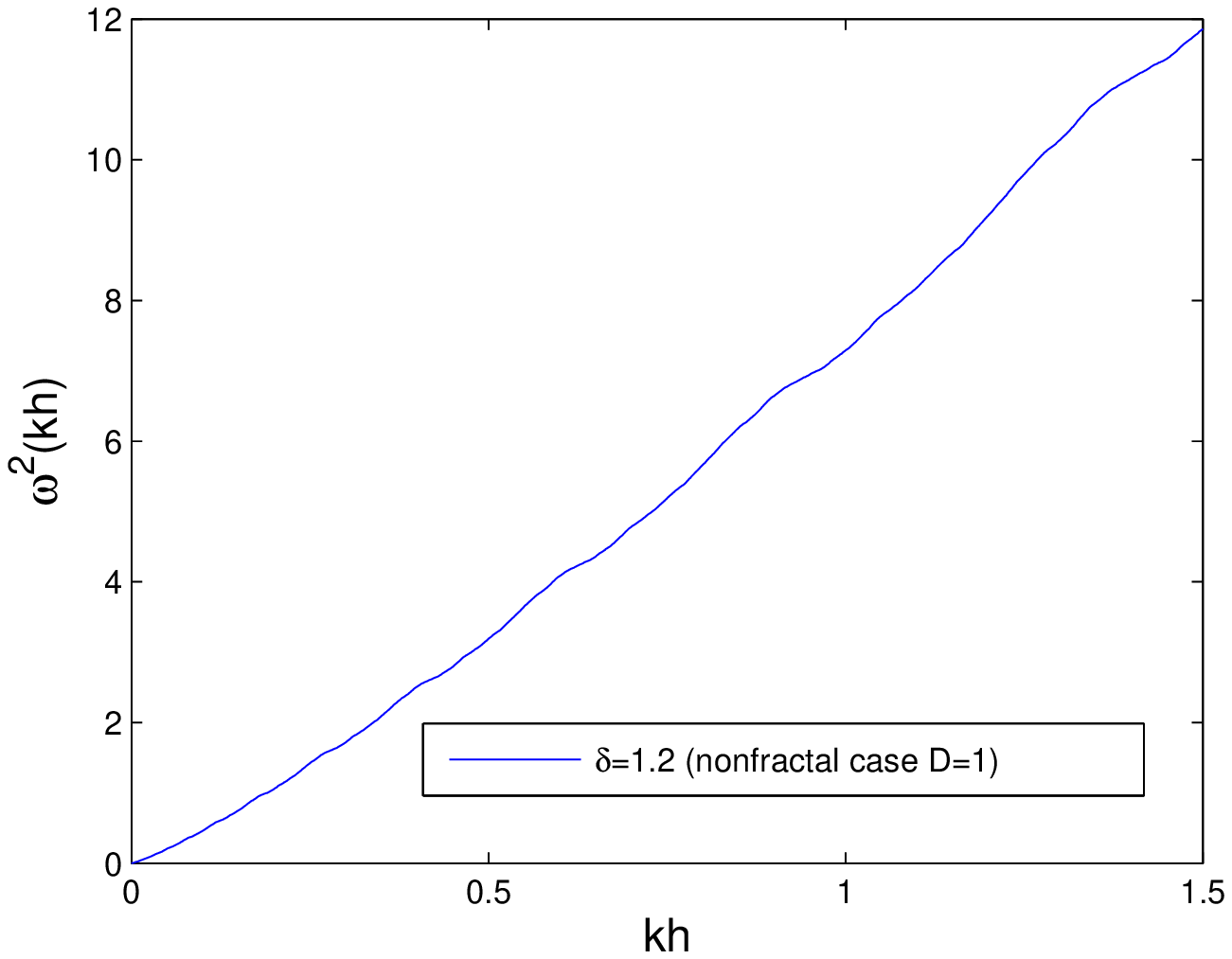,height=4in,width=4in}
\caption{Dispersion relation $\omega^2(kh)$ in arbitrary units for $N=1.5$ and $\delta=1.2$ }
\psfig{figure=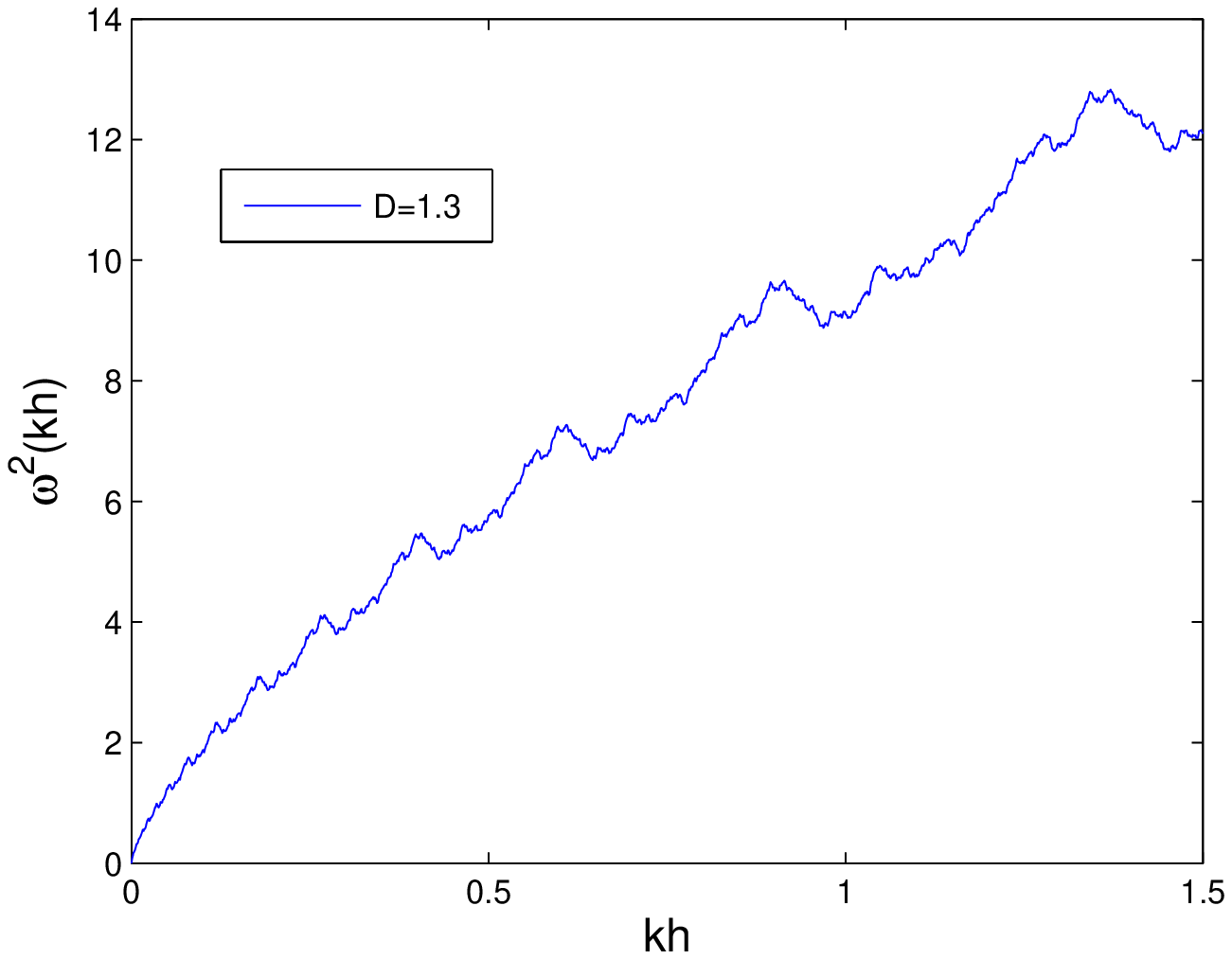,height=4in,width=4in}
\caption{Dispersion relation $\omega^2(kh)$ in arbitrary units for $N=1.5$ and $\delta=0.7$}
\end{figure}
\begin{figure}[t]
 \psfig{figure=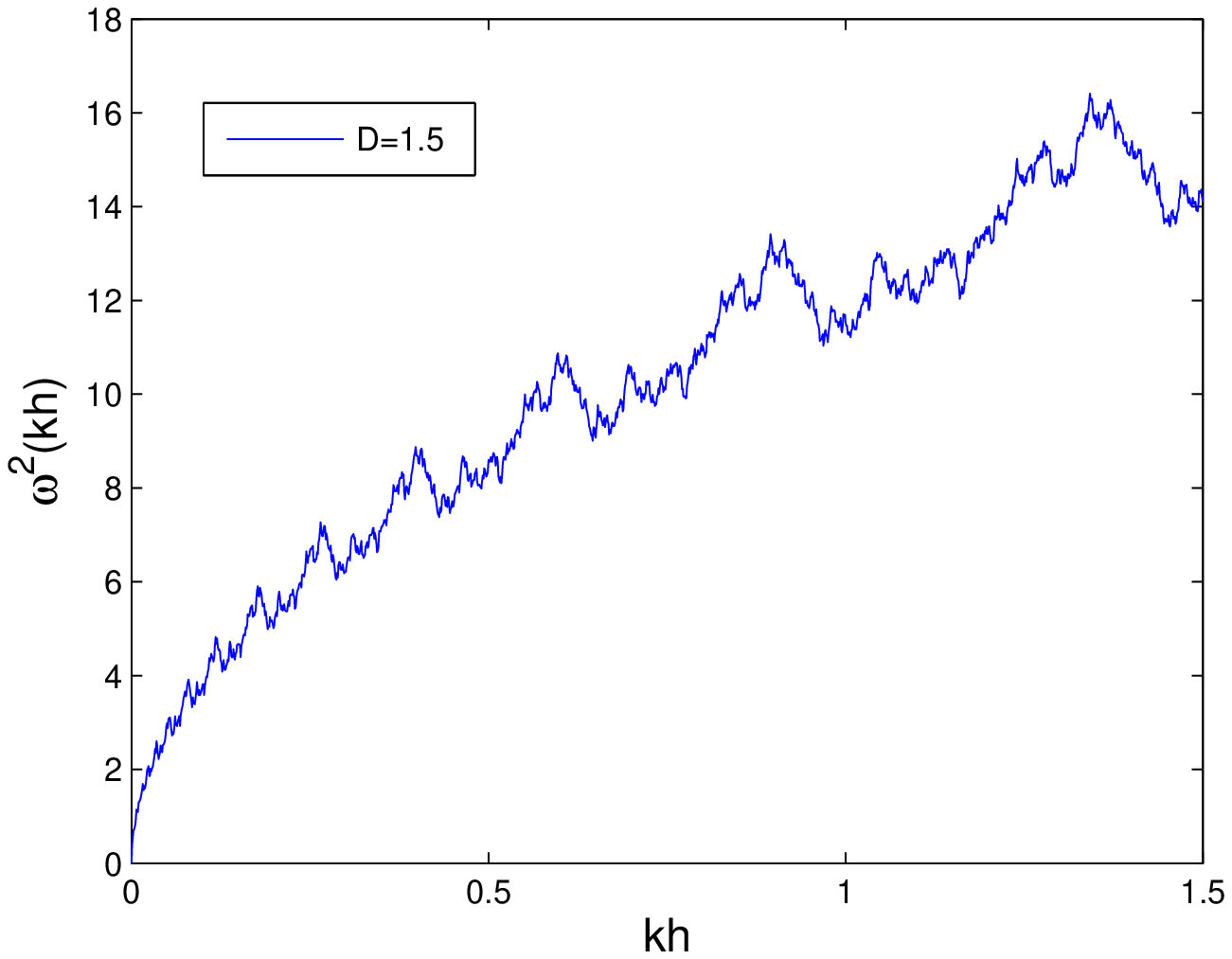,height=4in,width=4in}
 \caption{Dispersion relation $\omega^2(kh)$ in arbitrary units for $N=1.5$ and $\delta=0.5$}
\psfig{figure=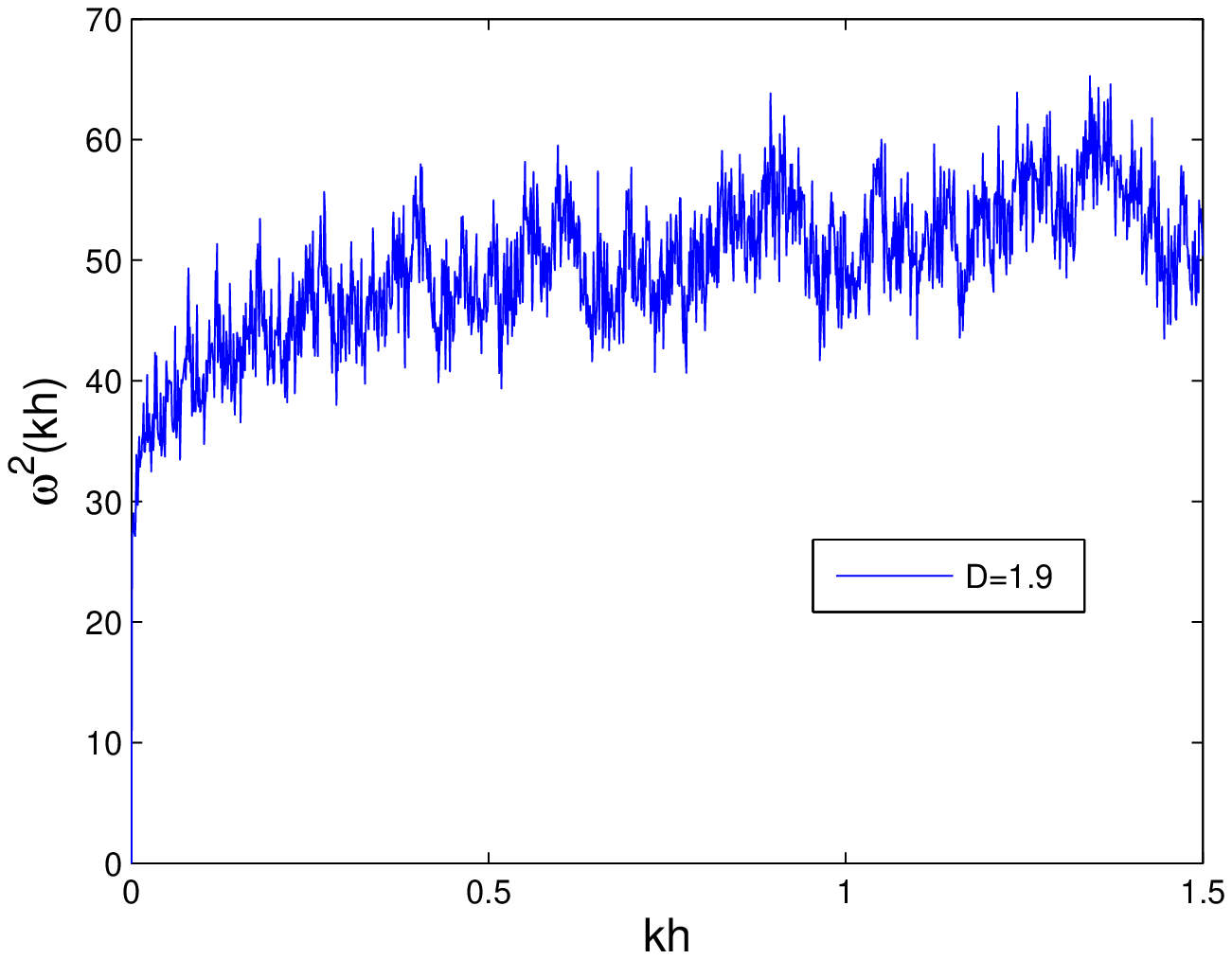,height=4in,width=4in}
\caption{Dispersion relation $\omega^2(kh)$ in arbitrary units for $N=1.5$ and $\delta=0.1$}
\end{figure}

\bibliographystyle{spmpsci}

\end{document}